# Metamorphic InAs/InGaAs QWs with electron mobilities exceeding $7\times10^5$ cm²/Vs

A. Benali,[1] P. Rajak,[1] R. Ciancio,[1] J.R. Plaisier,[2] S. Heun,[3] G. Biasiol[1*]

[1] *IOM CNR, Laboratorio TASC, Area Science Park Basovizza, Trieste, 34149, Italy*

[2] *Elettra-Sincrotrone Trieste S.C.p.A, Area Science Park Basovizza, Trieste, 34149, Italy*

[3] *NEST, Istituto Nanoscienze-CNR and Scuola Normale Superiore, Pisa, 56127, Italy*

We present a study on the influence of strain-relieving InAlAs buffer layers on metamorphic InAs/InGaAs quantum wells grown by molecular beam epitaxy on GaAs. Residual strain in the buffer layer, the InGaAs barrier and the InAs wells were assessed by X-ray diffraction and high-resolution transmission electron microscopy. By carefully choosing the composition profile and thicknesses of the buffer layer, virtually unstrained InGaAs barriers embedding an InAs quantum well with thickness up to 7nm can be grown. This allows reaching low-temperature electron mobilities much higher than previously reported for samples obtained by metamorphic growth on GaAs, and comparable to the values achieved for samples grown on InP substrates.

[*] Corresponding author

tel. +39 040 3756439

e-mail biasiol@iom.cnr.it

**Highlights:**

- We have achieved record-high electron mobilities for InAs/InGaAs QWs grown on GaAs
- Strain in the metamorphic QW can be tailored by optimization of the buffer layer
- Such optimization allows to increase the QW thickness from 4nm to up to 7nm
- Strain analysis was performed by synchrotron-radiation XRD and cross-sectional TEM

**Keywords:**

A1. High resolution X-ray diffraction, Stresses

A3. Molecular beam epitaxy, Quantum wells

B2. Semiconducting III-V materials, Semiconducting indium compounds

1. INTRODUCTION

InAs-based 2D electron gases (2DEGs) are potential platforms for a class of low-temperature applications taking advantage of strong spin-orbit coupling, large g-factor, and interface transparency to superconductors [1]. Unfortunately, the difficulty in finding lattice-matched substrates limits the flexibility of heterostructure engineering. While almost lattice-matched GaSb substrates allow growth of InAs quantum wells (QW) with thickness up to 24nm and low-T electron mobility µ ~ $2\times10^6$ cm$^2$/Vs [2], growth on the more technologically relevant and resistive InP and GaAs wafers limits the QW thickness to a few nm due to strain, with µ up to ~ $10^6$ cm$^2$/Vs and ~ $5\times10^5$ cm$^2$/Vs on InP and GaAs, respectively [1,3]. In this paper, we show that metamorphic InAs/InGaAs QWs can be achieved on GaAs with similar quality to InP. With respect to our previous work [3], optimization of strain relief allowed us to increase QW thickness from 4 to 7nm, thus decreasing interface and alloy scattering.

2. EXPERIMENT

InAs/In$_{0.81}$Ga$_{0.19}$As QWs were grown by solid source Molecular Beam Epitaxy (MBE) on semi-insulating (001) GaAs substrates. To ensure the formation of a metamorphic, dislocation-free QW region, we inserted a strain-relieving In$_x$Al$_{1-x}$As buffer layer (BL), by adapting a technique previously used for In$_{0.75}$Ga$_{0.25}$As QWs [4], to favour the accommodation of the QW lattice parameter on the GaAs substrate. The detailed layer sequence and growth conditions are described elsewhere [3,4]. The 2DEG region consists of a 7 nm-thick InAs QW embedded into 9 nm-thick In$_{0.81}$Ga$_{0.19}$As layers with In$_{0.81}$Al$_{0.19}$As barriers, placed at 120 nm from the surface (see Fig. 1a). Increasing the InAs QW thickness from 4 [3] to 7 nm allows to increase the fraction of the 2DEG density contained in the binary QW region from 45% to 69% (as can be seen in the 1D Poisson-Schrödinger simulation of Fig 1b and c [5]), thus reducing effects interface and alloy scattering on low-T electron mobility. Note that 9 nm In$_{0.81}$Ga$_{0.19}$As layers on both sides of the QW are enough to ensure that virtually the whole 2DEG is contained within the InAs/In$_{0.81}$Ga$_{0.19}$As sandwich. To sustain the additional strain created by making the InAs QW thicker, we increased the In composition of the InGaAs and InAlAs regions from 0.75 to 0.81. The 1.2 µm-thick BL consists of 50nm- thick In$_x$Al$_{1-x}$As steps with $x$ increasing from 0.15 to 0.81, terminated by a top In$_{0.84}$Al$_{0.16}$As step of thickness $t$ (Fig. 1, left). A series of samples were grown by varying $t$ from 50 to 400 nm in order to tune the strain in the x=0.81 barrier regions and in the QW.

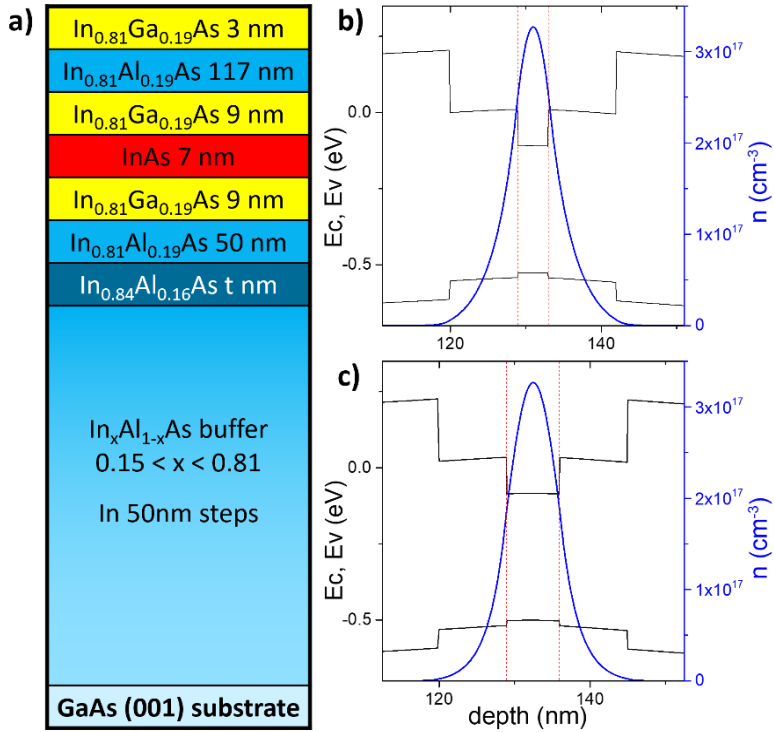

*Figure 1. (a) Schematics of the growth sequence. (b, c) 1D Poisson-Schrödinger simulations of a 4nm (b) and 7nm (c) InAs/In$_{0.84}$Ga$_{0.16}$As/In$_{0.84}$Al$_{0.16}$As QW [5]. Valence and conduction band profiles are indicated in black, while 3D electron density is indicated in blue. Boundaries of the InAs QWs are marked as vertical red dashed lines. A ~3×10$^{16}$ cm$^{-3}$ n-type background doping was assumed in the In$_{0.84}$Al$_{0.16}$As barriers [4].*

X-Ray Diffraction (XRD) experiments were carried out at the MCX beamline of the Elettra Synchrotron (Trieste, Italy) [6]. Measurements of the In concentration and the residual strain in the QW were carried out using a Huber high resolution X-ray diffractometer using an X-ray energy of 8keV. Symmetric (004) Bragg scans were used to measure the lattice parameter in the growth direction. The exact composition of the 0.81 and 0.84 regions was determined by asymmetric (224) Bragg scans of thick layers with complete lattice relaxation (not shown).

High-resolution Transmission Electron Microscopy (HRTEM) and analyses were performed in cross-section by using a JEOL 2010 UHR field emission gun microscope operated at 200 kV with a measured spherical aberration coefficient $C$s of 0.47 ± 0.01 mm. Cross-sectional TEM specimens were prepared using conventional mechanical polishing followed by dimpling, and ion etching with Ar-gas with angle 6º and accelerating voltage 3.5 kV. We estimated the strain in the InAs QWs by using geometric phase analysis (GPA) [7]. In the GPA method, any shift of the atomic lattice relative to some reference lattice in an image is determined by mapping corresponding local shifts of diffraction peaks in the Fourier transform of the image. The scripts implemented in the FRWR tools menu [8] for Gatan Digital Micrograph software was used to calculate the strain variation. [11$\bar{1}$] and [1$\bar{1}$1] Bragg reflections were used to calculate the 2D symmetric strain map within the QW. A circular aperture around Bragg spots was applied to measure the strain map with a spatial resolution of 3 nm.

Transport measurements were performed at 4.2 K on ~ 4×4 mm$^2$ van der Pauw structures.

3. RESULTS AND DISCUSSION

Fig. 2a shows (004) XRD rocking curves of five representative samples with different $t$. A broad feature at lower angles with respect to GaAs (004) comes from emission of the BL, with two distinct peaks emerging on the left-end side, corresponding to the $In_{0.81}Al_{0.19}As$ and $In_{0.84}Al_{0.16}As$ regions. Dashed lines indicate clear peak shifts for both $In_{0.81}Al_{0.19}As$ and $In_{0.84}Al_{0.16}As$ as $t$ increases. Note that within the broad BL feature, a peak emerges at about $\omega \approx 31.4°$. This peak is due to an accumulation of $In_xAl_{1-x}As$ steps with x around 0.6 in the design of the BL [9], and slightly shifts to the right with increasing $t$ similarly to the 0.81 and 0.84 regions, as a result of a different strain propagation within the BL. Fig. 2b shows the residual perpendicular strain in the x=0.81 and 0.84 regions as a function of $t$, calculated assuming constant compositions in all samples inferred from the (224) scans on thick layers. For *t* above 200nm, the 0.84 region becomes virtually strain-free, while in the 0.81 region the strain switches from compressive to tensile.

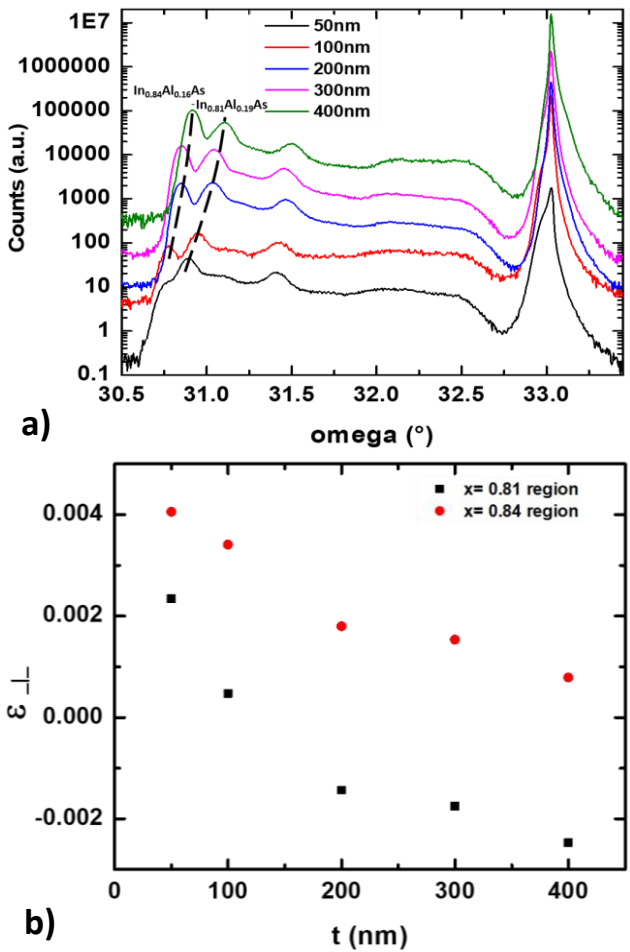

*Figure 2. (a) (004) XRD rocking curves showing GaAs and $In_xAl_{1-x}As$ Bragg peaks for different t. Dashed lines indicate the x=0.81 and 0.84 region peak shifts. (b) Residual perpendicular strain in the x=0.81 and 0.84 regions as a function of t.*

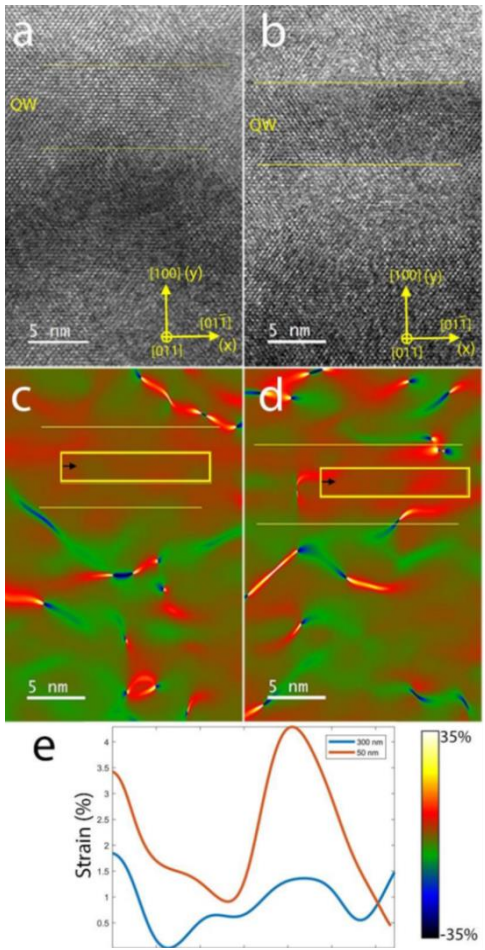

*Fig. 3: Cross-sectional HRTEM images of the QW region for (a) t = 300 and (b) 50 nm. Strain maps for (c) t = 300 and (d) 50 nm. (e) Strain profiles from the yellow boxes in (c) and (d) along black arrows.*

Figs. 3a and 3b represent cross sectional HRTEM images of the QW regions taken along [011] zone axis the for *t* = 300 nm and 50 nm, respectively. The QW interfaces are marked with yellow lines in the images. Both samples exhibit flat and atomically sharp interfaces. From the HRTEM images, the strain variation within the QW is calculated with respect to the $In_{0.81}Al_{0.19}As$ layers using GPA with <111> reflections. Out-of-plane lattice strain maps $\varepsilon_{yy}$ of QW for *t* = 300 nm and 50 nm are presented in Figs 3c and 3d, respectively. It can be noted that the QW region for *t* = 50 nm has a higher average strain and exhibits sharp variations, which for *t* = 300 nm are present in the barrier only. To quantify the strain within the QW, horizontal line scans have been taken from the region marked in the yellow box (Figs. 3c,d), averaged over 100 pixels along the growth direction (Fig. 3e). The calculated mean out-of-plane strain values within QW for *t* = 300 nm and 50 nm are 0.9±0.5 % and 2.2±1.1 % respectively.

Hall effect measurements at 4.2 K are shown in Fig. 4a. It can be seen that $\mu$ increases from $6\times10^4 cm^2/Vs$ for *t*=50nm up to $7.1\times10^5 cm^2/Vs$ for *t*=300nm, i.e., more than an order of magnitude. For all the samples, *n* lays in the $3-3.5\times10^{11}$ $cm^{-2}$ range, largely independent of *t*. Shubnikov-de Haas oscillations for *t*=300nm confirm the formation of a 2DEG without parasitic conduction channels, and the onset of integer quantum Hall plateaus (Fig. 4b). Mobility data of Fig. 4a, together with the XRD and TEM analyses, indicate that strain plays a major role as a scattering mechanism limiting electron mobility in metamorphic InAs QWs, and that a careful design of the strain-relieving $In_xAl_{1-x}As$ BL can increase electron mobility by more than one order of magnitude. The most substantial mobility variations take place for *t* in the ~ 50 – 300 nm range, where strain changes in the barriers (and thus likely in the InAs QW too) are more marked; for larger *t* both residual strain and mobility tend to saturate. Note that, although presence of dislocations in the InAs QW could be hinted by cross-sectional TEM for *t* = 50nm (Fig. 3d), they are not likely to be the limiting scattering mechanisms in our lower-mobility samples, as the onset of a significant network of misfit dislocations [10] is accompanied by a sharp drop of mobility (rather than a continuous decrease as in our case) and a considerable increase of the electron density (that we do not observe). Likewise, such dislocations cause the appearance of surface defects visible by atomic force microscopy as deep trenches [9,10], while on our set of samples the surface morphology is largely independent on *t*, and presents in all of them a similar cross-hatched corrugation pattern typical for InGaAs/GaAs metamorphic growth [9], with no evidence of defects and similar RMS roughness.

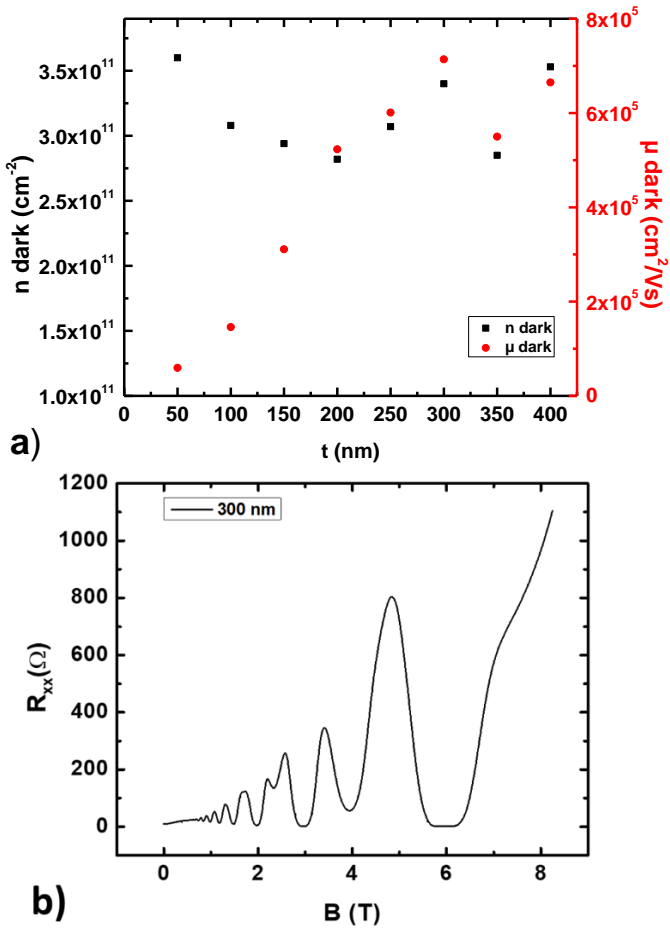

*Figure 4. (a) Low-temperature (T=4.2 K) electron charge density and mobility in the InAs/In$_{0.81}$Ga$_{0.19}$As 2DEG as a function of t. (b) Longitudinal resistance R$_{xx}$ as a function of magnetic field B for t = 300nm.*

4. CONCLUSIONS

We have shown that proper design of the strain-relieving BL strongly influences residual strain and electron mobility of metamorphic InAs/In$_{0.81}$Ga$_{0.19}$As QWs. In particular, terminating the BL with an In$_{0.84}$Al$_{0.16}$As step with at least 300 nm thickness minimizes such strain and results in low-temperature electron mobilities of about $7.1 \times 10^5$ cm$^2$/Vs at *n* ~ $3 \times 10^{11}$ cm$^{-2}$. This represents an increase by about a factor of 2 with respect to similar 4nm-thick InAs/In$_{0.75}$Ga$_{0.25}$As QW for the same electron densities [3]. The possibility to increase QW thickness and In composition in the barriers is expected to decrease interface and alloy scattering, resulting in an overall mobility increase. Further investigations of mobility dependence on electron density in gated Hall bars are being performed to elucidate the scattering mechanisms. Finally, our samples are competitive with state-of-the-art equivalent systems grown on InP substrates in terms of electron mobility for similar electron densities [1], providing the possibility of an alternative technological platform for applications including topological quantum computing.

## 5. ACKNOWLEDGMENTS

We acknowledge support from EU H2020 under the FETOPEN Grant Agreement No. 828948 (AndQC). P.R. acknowledges the receipt of a fellowship from the ICTP Programme for Training and Research in Italian Laboratories, Trieste, Italy.

REFERENCES

[1] A.T. Hatke, T. Wang, C. Thomas, G.C. Gardner, M.J. Manfra, Mobility in excess of $10^6$ cm$^2$/V s in InAs quantum wells grown on lattice mismatched InP substrates, Appl Phys Lett. 111 (2017). https://doi.org/10.1063/1.4993784.

[2] T. Tschirky, S. Mueller, Ch.A. Lehner, S. Fält, T. Ihn, K. Ensslin, W. Wegscheider, Scattering mechanisms of highest-mobility InAs/Al$_x$Ga$_{1-x}$Sb quantum wells, Phys. Rev. B. 95 (2017) 115304. https://doi.org/10.1103/PhysRevB.95.115304.

[3] D. Ercolani, G. Biasiol, E. Cancellieri, M. Rosini, C. Jacoboni, F. Carillo, S. Heun, L. Sorba, F. Nolting, Transport anisotropy in In$_{0.75}$Ga$_{0.25}$As two-dimensional electron gases induced by indium concentration modulation, Phys. Rev. B. 77 (2008) 235307. https://doi.org/10.1103/PhysRevB.77.235307.

[4] F. Capotondi, G. Biasiol, I. Vobornik, L. Sorba, F. Giazotto, A. Cavallini, B. Fraboni, Two-dimensional electron gas formation in undoped In$_{0.75}$Ga$_{0.25}$As/In$_{0.75}$Al$_{0.25}$As quantum wells, J. Vac. Sci. Technol. B Microelectron. Nanometer Struct. 22 (2004) 702. https://doi.org/10.1116/1.1688345.

[5] I. Tan, G.L. Snider, L.D. Chang, E.L. Hu, A self-consistent solution of Schrödinger–Poisson equations using a nonuniform mesh, J. Appl. Phys. 68 (1990) 4071–4076. https://doi.org/10.1063/1.346245.

[6] L. Rebuffi, J.R. Plaisier, M. Abdellatief, A. Lausi, P. Scardi, MCX: a Synchrotron Radiation Beamline for X-ray Diffraction Line Profile Analysis, Z. Für Anorg. Allg. Chem. 640 (2014). https://doi.org/10.1002/zaac.201400163.

[7] M.J. Hÿtch, E. Snoeck, R. Kilaas, Quantitative measurement of displacement and strain fields from HREM micrographs, Ultramicroscopy. 74 (1998). https://doi.org/10.1016/S0304-3991(98)00035-7.

[8] C. Koch, Determination of core structure periodicity and point defect density along dislocations, Ph.D. thesis, 2002.

[9] F. Capotondi, G. Biasiol, D. Ercolani, V. Grillo, E. Carlino, F. Romanato, L. Sorba, Strain induced effects on the transport properties of metamorphic InAlAs/InGaAs quantum wells, Thin Solid Films. 484 (2005). https://doi.org/10.1016/j.tsf.2005.02.013.

[10] F. Capotondi, G. Biasiol, D. Ercolani, L. Sorba, Scattering mechanisms in undoped $In_{0.75}Ga_{0.25}As/In_{0.75}Al_{0.25}As$ two-dimensional electron gases, J. Cryst. Growth. 278 (2005). https://doi.org/10.1016/j.jcrysgro.2004.12.104.